\documentclass[9pt,twocolumn,twoside,lineno]{pnas-new}

\usepackage{bm}

\templatetype{pnasresearcharticle} 

\title{Internal feedback in the cortical perception-action loop enables fast and accurate behavior}

\author[a,1]{Jing Shuang (Lisa) Li}
\author[a,b,1]{Anish A. Sarma} 
\author[c,d]{Terrence J. Sejnowski}
\author[a]{John C. Doyle}

\affil[a]{Control and Dynamical Systems, Division of Engineering and Applied Science, California Institute of Technology, Pasadena, CA 91125}
\affil[b]{School of Medicine, Vanderbilt University, TN 37232}
\affil[c]{Computational Neurobiology Laboratory, The Salk Institute for Biological Studies, La Jolla, CA 92037}
\affil[d]{Division of Biological Sciences, University of California San Diego, La Jolla, CA 92093}

\leadauthor{Li* \& Sarma*} 

\significancestatement{Internal feedback projections -- action output signals flowing from motor areas back to early sensory processing regions such as visual and auditory areas -- are ubiquitous and more numerous in sensorimotor structures than feedforward projections. However, the function of internal feedback is poorly understood, particularly in the context of task performance. We leverage control theory and simple models to demonstrate that internal feedback is necessary to facilitate good task performance when there are communication limitations such as speed-accuracy trade-offs. These trade-offs necessitate compensatory signals for self-generated and predictable movements. Control theory explains why motor-related signals are found throughout the cortex and why motor cortex is dominated by internal dynamics.}

\authorcontributions{J.S.L., A.A.S., T.J.S., and J.C.D. designed research. J.S.L. and A.A.S. performed research and analysis. J.S.L., A.A.S., T.J.S., and J.C.D wrote the paper.} 
\authordeclaration{The authors declare no competing interest.}
\equalauthors{\textsuperscript{1} J.S.L. and A.A.S. contributed equally to this work.}
\correspondingauthor{E-mails: jsli@caltech.edu, anish.a.sarma@vanderbilt.edu, terry@salk.edu, doyle@caltech.edu}


\begin{abstract}
Animals move smoothly and reliably in unpredictable environments. Models of sensorimotor control have assumed that sensory information from the environment leads to actions, which then act back on the environment, creating a single, unidirectional perception-action loop. This loop contains internal delays in sensory and motor pathways, which can lead to unstable control. We show here that these delays can be compensated by internal feedback signals that flow backwards, from motor towards sensory areas. Internal feedback is ubiquitous in neural sensorimotor systems and recent advances in control theory show how internal feedback compensates internal delays. This is accomplished by filtering out self-generated and other predictable changes in early sensory areas so that unpredicted, actionable information can be rapidly transmitted toward action by the fastest components. For example, fast, giant neurons are necessarily less accurate than smaller neurons, but they are crucial for fast and accurate behavior. We use a mathematically tractable control model to show that internal feedback has an indispensable role in achieving state estimation, localization of function -- how different parts of cortex control different parts of the body -- and attention, all of which are crucial for effective sensorimotor control. This control model can explain anatomical, physiological and behavioral observations, including motor signals in visual cortex, heterogeneous kinetics of sensory receptors and the presence of giant Betz cells in motor cortex,  Meynert cells in visual cortex and giant von Economo cells in the prefrontal cortex of humans as well as internal feedback patterns and unexplained heterogeneity in other neural systems. 
\end{abstract}

\dates{This manuscript was compiled on \today}
\doi{\url{www.pnas.org/cgi/doi/10.1073/pnas.XXXXXXXXXX}}

\begin{document}

\maketitle
\thispagestyle{firststyle}
\ifthenelse{\boolean{shortarticle}}{\ifthenelse{\boolean{singlecolumn}}{\abscontentformatted}{\abscontent}}{}

\begin{figure}
    \centering
    \includegraphics[page=2,width=0.8\linewidth]{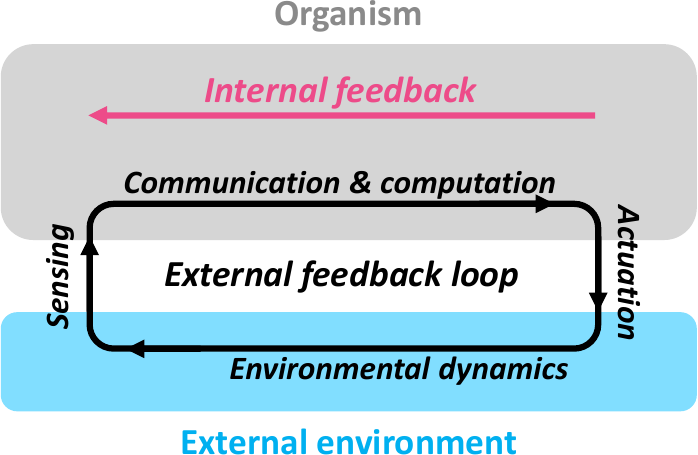}
    \caption{Single-loop model of sensorimotor control. The organism receives information from the external environment via sensors, communicates this information through the body, computes actions, then acts on the environment; this forms the \textit{external feedback loop}, or single loop model (black). Internal signals that flow opposite to the direction of the external feedback loop are classified as \textit{internal feedback} (pink). Thus, the internal feedback is \textit{counterdirectional}. Internal feedback also includes \textit{lateral interactions} within or between areas (not shown).}
    \label{fig:single_loop}
\end{figure}

\dropcap{F}eedback control is an essential strategy for both engineered and biological systems to achieve reliable movements in unpredictable environments \cite{doyle_feedback_1992}. Optimal and robust control theory, which provide a general mathematical foundation to study feedback systems, have been used successfully to explain behavioral observations by modeling the sensorimotor system as a single control loop called the perception-action cycle \citep{Wolpert1995,Todorov2004, Franklin2011}. In these models, the sensorimotor system senses the environment, communicates signals from sensors to the brain, computes actions, and then acts on the environment, feeding back to the sensors and forming a single unidirectional loop as shown in Fig. \ref{fig:single_loop}.

\begin{figure}
    \centering\includegraphics[page=1,width=0.9\linewidth]{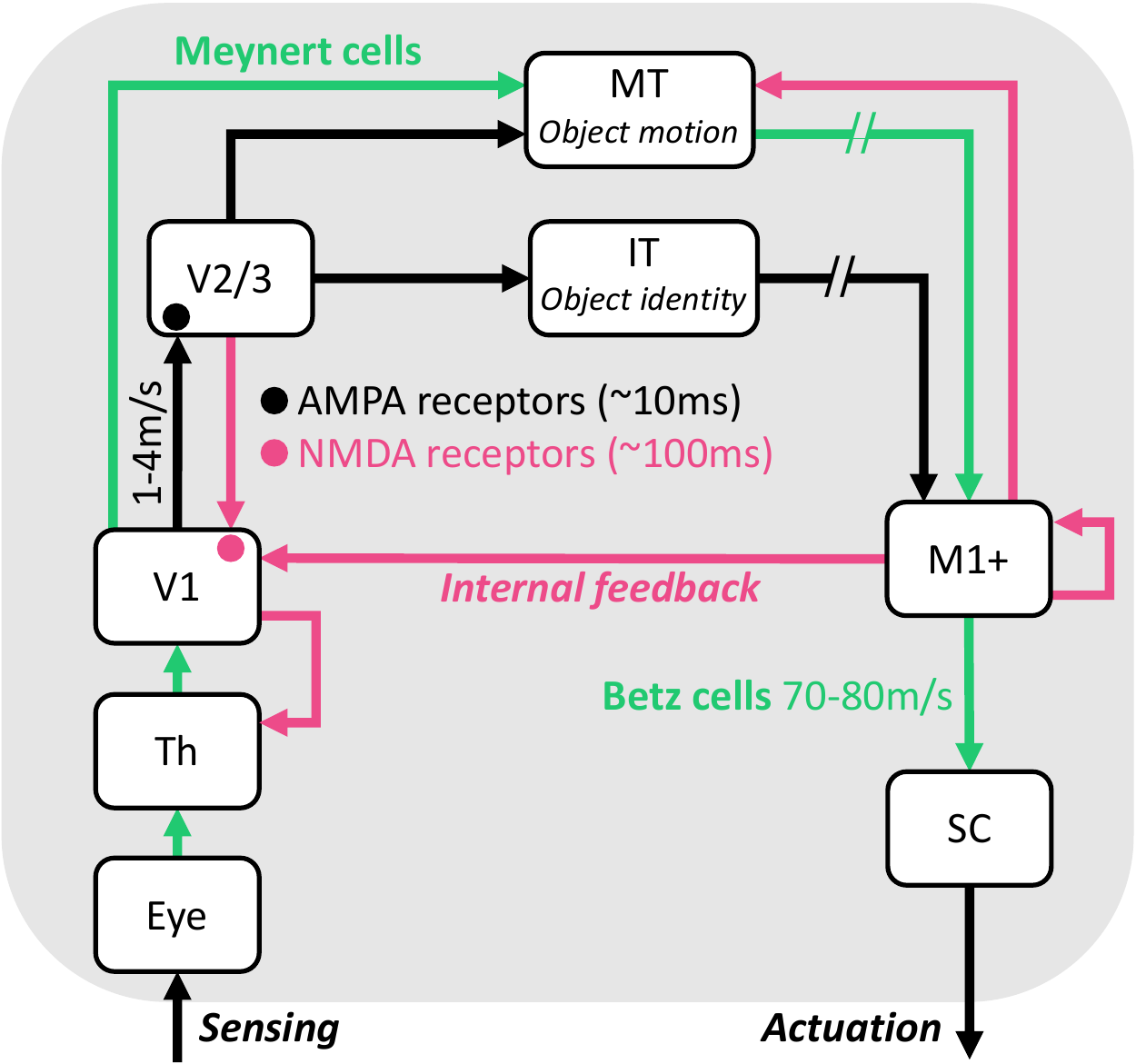}
    \caption{A partial, simplified schematic of sensorimotor control. We focus on key cortical and subcortical areas and communications between them. Black and green arrows indicate communications that traverse from sensing toward actuation; green arrows are particularly fast pathways, which enable tracking moving objects in our model. Pink arrows indicate internal feedback signals, which traverse from actuation toward sensing. Broken lines are not necessarily direct neuronal projections. SC = spinal cord, Th = thalamus, V1 = primary visual cortex, M1+ = primary motor cortex and additional motor areas, V2/3 = secondary and tertiary visual cortex, IT = inferotemporal cortex, MT = mediotemporal cortex (V5). Only a subset of the internal feedback pathways are shown (e.g. not included are internal feedback signals from M1+ to V2 and signals from M1+ to IT).
    \label{fig:intro_figure_neuro}} 
\end{figure}

Consider the canonical model of localized function in the primate visuomotor cortical pathway, depicted in Fig. \ref{fig:intro_figure_neuro}: A visual signal is encoded on the retina, then travels to the lateral geniculate nucleus (LGN) of the thalamus, and on to the primary visual cortex (V1), progressing through successive transformations until it reaches the primary motor cortex (M1), the spinal cord, and ultimately the muscles. The single-loop feedback model also makes implicit assumptions about the interpretation of responses from sensory and motor populations of neurons, which represent sensory signals and action signals, respectively. Although intuitive, this model neglects a well-known and ubiquitous feature of sensorimotor processing: \textit{internal feedback}, which is the main focus of this paper \cite{zagha2020}.

The perception-action control model does not have a direct role for internal feedback connections. Internal feedback includes all signals that do not flow from sensing towards action. We can divide internal feedback into two broad categories: \textit{counterdirectional} between brain areas and \textit{lateral interactions} within or between areas. Counterdirectional internal feedback is in the opposite direction of the single-loop model (for instance, from V2 to V1); these signals flow from action toward sensing. Lateral internal feedback consists of recurrent connections within and between areas (for instance, from V2 to V2, or from MT to IT). This distinction emphasizes the importance of where control signals are spatially located. 

One reason that the single-loop model has endured is that it offers a set of tools from control theory and a conceptual framework that allows  subsystems to be treated in isolation. However, these subsystems are not isolated, and with internal feedback each subsystem has access to both bottom up and top down information. The eye is itself a site of computation and control: as the eye moves and senses different parts of the visual scene, lateral interactions within the retina control spatial and temporal filter properties that can adapt and identify important features under a wide range of illumination and scene dynamics \cite{Zhaoping2014,GollischMeister2010}. Retinal ganglion cells project to relay neurons in the LGN, which then project to primary visual cortex, V1, but a much greater number of feedback neurons project back from V1 to LGN \cite{Callaway2004,Angelucci2003,ElShamaylehKumbhani2013}. 

Projections from motor areas in cortex to visual areas have a wide range of morphology, myelination and synapse kinetics \citep{Felleman1991, Callaway2004, Muckli2013}.
Given the position of M1 in the final common pathway, one might expect activity in M1 to be driven by current visual stimuli or current movements, but instead autonomous or top-down preparatory activity with internal rotational dynamics dominate the data \cite{Churchland2012}. Counterintuitively, signals related to movements of the whole body are found in areas typically associated with particular parts of the body, such as the hand area, as well as sensory areas such as primary visual cortex \cite{Willett2020,Stavisky2019, Stringer2019, Musall2019}. Indeed, recent analysis of the correlation structure between neurons during a visual discrimination task revealed a task-related global mode in the correlations between cortical neurons associated with the task response rather than the sensory stimulus, strongly supporting the idea that top-down feedback is an important element of sensory processing \cite{Ebrahimi2022}. 
Although not typically studied together, all of these sensory and motor signals are generated by internal feedback and are the focus of this study.

Internal feedback has been studied in the context of predictive coding \citep{RaoBallard1999,Keller2018} and are invoked in other models \citep{VyasShenoy2020, KarDiCarlo2019, BastosFriston2012,Swain07}. However, these models focus on sensory or motor systems separately and do not account for key constraints on neuronal communication in both space and time to achieve sensorimotor tasks. Achieving fast and accurate computation and communication over large distances is difficult and often impossible because communication may be slow, limited in bandwidth and constrained to spatially localized populations. 

Here, we build on the foundations of recent work in distributed control theory \cite{Anderson2019,Nakahira2021,Paper1,Paper2,Paper3} and show that internal feedback is a solution to achieving rapid and accurate control given the spatial and temporal constraints on brain components and communication systems. We analyze an idealized class of control models and prove mathematically that internal feedback is both plausible and \textit{necessary} for achieving optimal performance.
Internal feedback serves at least three functions in our model: \textit{state estimation}, \textit{localization of function} and \textit{focused attention}, all of which are crucial for effective sensorimotor control and survival. This theory explains why there are differences in population responses between M1 and V1, why different projections predominately activate AMPA or NMDA glutamate receptors, the functions of giant pyramidal cells in visuomotor control, and both the uses and limitations of localization of function in cortex. There is a general principle behind all of these physiological properties.

\subsection*{Task model and performance}
We analyze expected values and theoretical bounds on task performance for models of a simple, well-studied and ethologically relevant \textit{tracking} task, consisting of reaching for and grasping a moving object. The goal of the task is continuous pursuit, rather than one-time contact between limb and object.

The complete tracking task requires identification of the object in a cluttered visual scene, prediction of the object's movement, and generation and execution of bimanual limb movement. We make simplifying assumptions that are not essential to our conclusions, but allow us to study internal feedback in an accessible way using familiar linear dynamical systems. 

\subsubsection*{Single-loop feedback control}
Consider the task of tracking a moving object with the endpoint of a limb on a plane. The variable to be controlled is the \textit{tracking error} -- the distance between the hand and the object. We start by assuming that the system controlling the limb can perfectly sense the position of limb and object at every instant, which will be relaxed in later models. The \textit{cost} is defined as the Euclidean norm of the tracking error over time, with smaller cost indicating better tracking.

Let $x$, $u$, and $w$ represent the tracking error, the control action on the limb, and the action of the object, respectively. We will refer to $x$ as the \textit{state} of the system. Let $A$ be a matrix that represents the intrinsic dynamics of $x$ -- including the mechanical coupling between two dimensions of limb movement. The time-evolving dynamics of the tracking error follows from a linear equation of motion:

\begin{equation} \label{eq:standard_system}
    x(t+1) = Ax(t) + u(t) + w(t)
\end{equation}

Let $\alpha$ denote the magnitude of the maximum eigenvalue of $A$, as a proxy for task difficulty. Note that $\alpha < 1$ corresponds to a task in which tracking error $x$ will decrease with no limb action, an easy task.

The actions $u$ provide feedback control on the tracking error, computed by an arbitrary function $\bm{K}$ that has access to all past and present tracking errors $x(1:t)$, as follows:

\begin{equation}
    u(t) = \bm{K}(x(1:t))
\end{equation}

The optimal solution to this problem is the linear quadratic regulator (LQR) and the optimal controller is $\bm{K}(x(1:t)) = -Ax(t)$ \cite{doyle_feedback_1992}. This controller fits into the single-loop model of sensorimotor control, as there is no internal feedback, and the addition of internal feedback does not provide any additional performance advantage. 

Controllers without internal feedback are optimal for a large but special class of problems, including classical state feedback and full control problems from control theory. Though mathematically elegant, these problems make assumptions that are impractical when applied to biological systems. In subsequent sections, we relax some of the assumptions implicit in this single-loop model and show that small deviations from assumptions relevant to biological systems give rise to the necessity of introducing internal feedback from which advantages accrue.

Any of the controllers in subsequent sections can be implemented in a variety of ways, though whether or not a particular controller needs internal feedback is generic across all possible implementations. In order to study internal feedback, we choose particular non-unique controller implementations which, when optimized, attain the optimal performance over the relevant class of controllers. We choose implementations for which the optimal solution is relatively transparent and easy to interpret.

\subsection*{State estimation requires counterdirectional internal feedback}

\subsubsection*{Internal feedback facilitates implicit estimation in the presence of sensor delays} 
\begin{figure}
    \centering
    \includegraphics[page=3,width=0.8\linewidth]{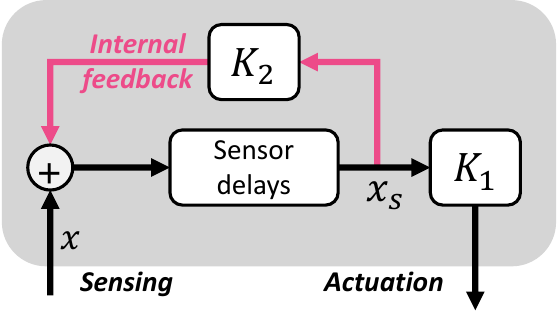}
    \caption{Optimal control model for system with sensor delays. Tracking error $x$ is sensed, then communicated by the sensor with some delay to the $K_1$ block, which computes the appropriate actuation. Counterdirectional internal feedback (pink) conveys information from actuation back toward sensing. Internal computation $K_2$ adjusts the sensor signal to compensate for actions taken by the system; this results in improved performance.}
    \label{fig:sensing_delay}
\end{figure}

Simple modifications to the control problem described above lead to an optimal controller $K$ whose implementation requires internal feedback. One such modification is the introduction of sensor delays, which are ubiquitous in biological systems (for instance, the neuronal conduction time from the eye to motor cortex, on the order of tens or hundreds of milliseconds). Sensor delays can be modeled by introducing a virtual internal state $x_s$, which represents the adjusted tracking error from the previous time step \cite{Paper2}. This formulation allows us to pose the delayed-sensor tracking problem as a standard control problem which can be optimally solved by LQR:
\begin{equation}
    \begin{gathered}
    \tilde{A} =\begin{bmatrix} A & 0 \\ I & 0\end{bmatrix}, C = \begin{bmatrix}0& I \end{bmatrix}
    \\\begin{bmatrix}x(t+1)\\x_s(t+1)\end{bmatrix} =\tilde{A}\begin{bmatrix}x(t)\\x_s(t)\end{bmatrix}+ u(t) + \begin{bmatrix} w(t) \\ 0 \end{bmatrix}
    \\u(t) = KC\begin{bmatrix}x(t)\\x_s(t)\end{bmatrix}
    \end{gathered}
\end{equation}
where the virtual internal state $x_s$ contains delayed information about the tracking error.
Controller $K$ can be partitioned into two block-matrices, ($K = \begin{bmatrix} K_1^\top & K_2^\top \end{bmatrix}^\top$). The resulting system is shown in Fig. \ref{fig:sensing_delay}. Here, the controller does not directly "perceive" the tracking error $x$ and only has access to the virtual internal state $x_s$. However, the controller can freely take actions that affect both the tracking error and the virtual state. The action on the virtual state, as shown in Fig. \ref{fig:sensing_delay}, is an example of counterdirectional internal feedback with gain $K_2$.

\begin{figure}
    \centering
    \includegraphics[page=8,width=0.7\linewidth]{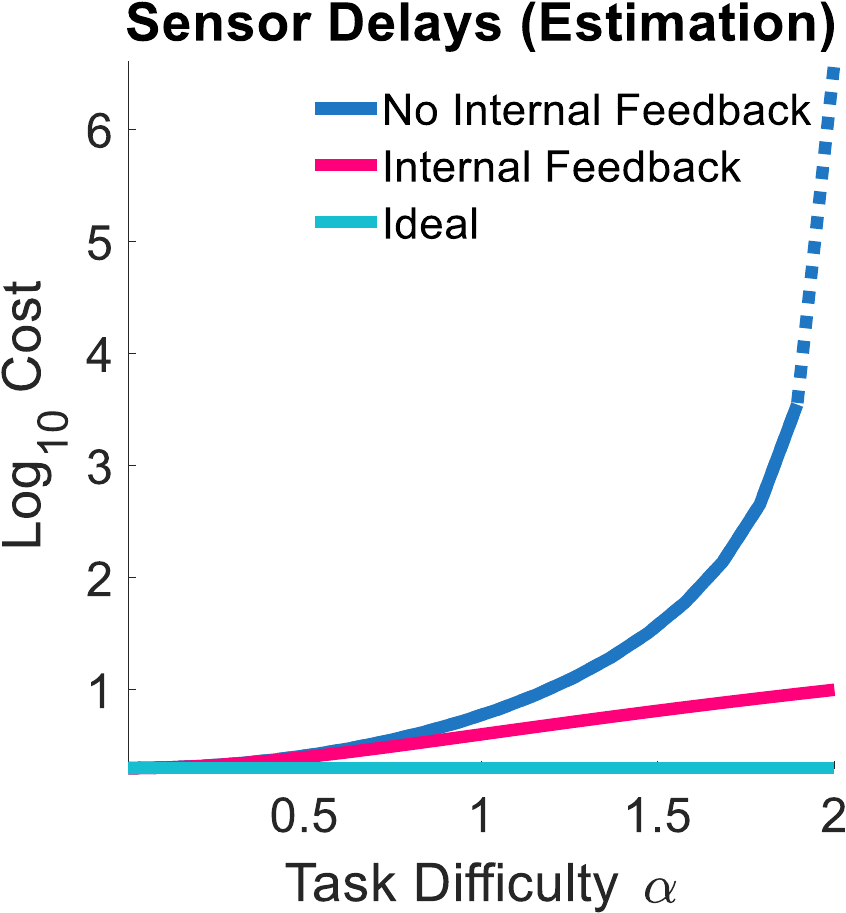}
    \caption{Delays in simple and otherwise unconstrained sensing necessitate internal feedback for good performance. In simulations, we consider the scalar problem of tracking a moving target over a line, varying the task difficulty (dynamics $A =  \alpha$). The `Ideal' controller contains no sensor delays. The `Internal Feedback' controller contains sensor delays, and uses internal feedback to compensate; the resulting performance is good (low-cost). The `No Internal Feedback' controller contains sensor delays, but uses no internal feedback; the resulting performance is poor (high cost). As $\alpha$ approaches 2, the task becomes infeasible without internal feedback (broken line).}
    \label{fig:fc_ifp}
\end{figure}

For the delayed sensing problem, the optimal controller has a simple analytical form: $K_1 = -A^2$ and $K_2 = -A$ is the internal feedback. If no internal feedback is allowed (i.e. we enforce $K_2 = 0$), then the optimal controller is $K_1 = -A^2/4$. We compare the performance of these two controllers in Fig. \ref{fig:fc_ifp}, and see that the controller with internal feedback far outperforms the controller without internal feedback. We also note that as the task becomes more difficult, the controller without internal feedback is unable to stabilize the closed-loop system and tracking breaks down.

For a controller with sensory delays, internal feedback is required for optimal performance. This also applies to controllers with actuator delays \cite{Paper2}. In both cases, internal feedback adjusts delayed signals to compensate for actions taken and information received during the delay; in other words, internal feedback implicitly compensates for the delays.

\subsubsection*{Intrinsic internal feedback in the Kalman filter}
\begin{figure}
\centering
\includegraphics[page=4,width=\linewidth]{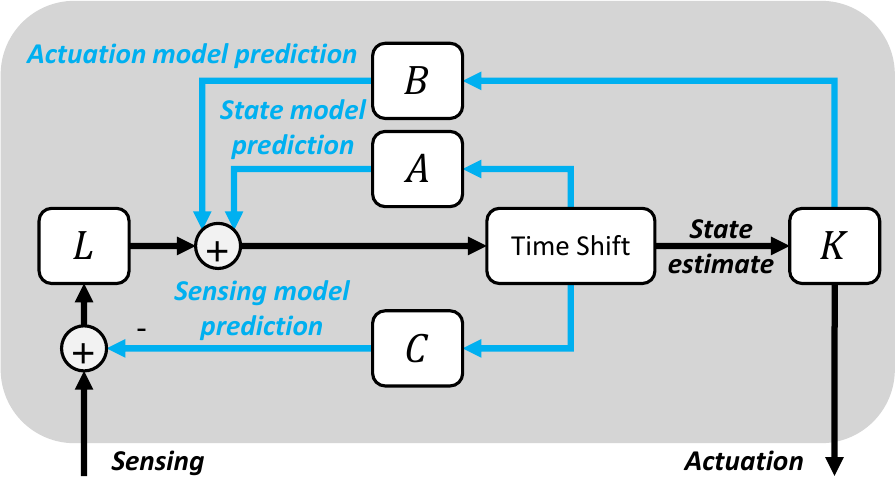}
\caption{Internal feedback in a controller with instantaneous but imperfect sensing and actuation. $A$, $B$, and $C$ represent the state, actuation, and sensing matrices of the physical plant; $K$ represents the optimal controller, and $L$ represents the optimal observer. The Time Shift block shifts $\hat{x}(t+1)$ to $\hat{x}(t)$ in Eq. 5. The internal feedback pathways (blue) are inherent to the Kalman Filter; these use state, actuation, and sensing models to create an internal estimate of the tracking error, or state. All internal feedback depicted in this diagram is counterdirectional.}
\label{fig:of_standard_ifp}
\end{figure}

We now consider the case in which sensing is instantaneous, but imperfect. Consider the following system:
\begin{equation}
\begin{gathered}
    x(t+1) = Ax(t) + Bu(t) + w(t) \\
    y(t) = Cx(t)
\end{gathered}
\end{equation}
where $y$ is the sensor input. Matrix $B$ represents the effect of action $u$ on tracking error $x$, and matrix $C$ represents how sensor input $y$ is related to tracking error $x$. This is a standard formulation in control theory, and the optimal controller makes use of controller gain $K$ and estimator gain $L$ as follows:
\begin{equation}
\begin{gathered}
    \hat{x}(t+1) = A\hat{x}(t) + Bu(t) + L(y(t) - C\hat{x}(t)) \\
    u(t) = K\hat{x}(t)
\end{gathered}
\end{equation}
where $\hat{x}$ is an internal estimate of tracking error $x$. This optimal controller uses the Kalman filter, which inherently contains three counterdirectional internal feedback pathways irrespective of delays being present. These pathways are represented by the blue arrows through in Fig. \ref{fig:of_standard_ifp}, and play a central role in state estimation. The pathway through $A$ estimates state evolution in the absence of noise and actuation; the pathway through $B$ accounts for controller action, and the pathway through $C$ predicts incoming sensory signals based on the internal estimated state.

\begin{figure}
\centering
\includegraphics[page=5,width=\linewidth]{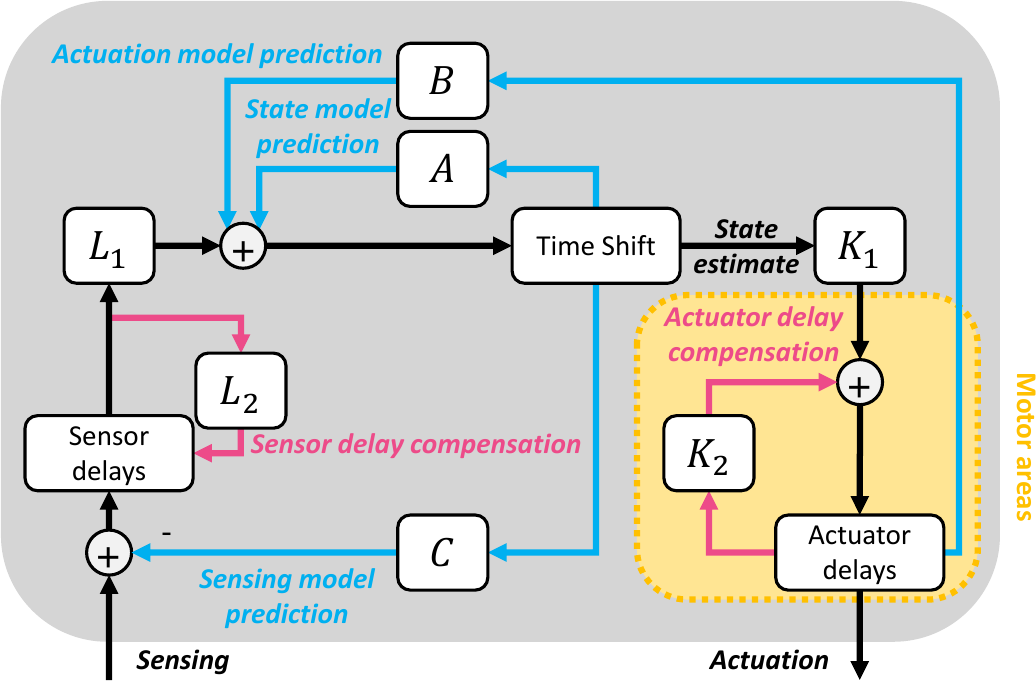}
\caption{Internal feedback in a controller with sensor and actuator delays. $A$, $B$, and $C$ represent the state, actuation, and sensing matrices of the physical plant; $K_1, K_2, L_1, L_2$ are submatrices of the optimal controller and observer gains. The internal feedback pathways (pink) through $L_2$ and $K_2$ compensate for sensor and actuator delays, respectively. Other internal feedback pathways (blue) are inherent to the Kalman Filter. All internal feedback depicted in this diagram is counterdirectional. The yellow box contains parts of the controller that roughly correspond to motor areas in cortex.}
\label{fig:of_delay_ifp}
\end{figure}

\subsubsection*{Sources of internal feedback are preserved in a Kalman filter with delays}
We now create a model that combines features from previous sections: sensor delays, actuator delays, and imperfect sensing. The problem can be written using virtual states as follows:
\begin{equation}
\begin{gathered}
\begin{bmatrix} x(t+1) \\ x_a(t+1) \\ x_s(t+1) \end{bmatrix} = 
\begin{bmatrix} 
        A & {B} & 0 \\
        0 & 0 & 0 \\
        {C} & 0 & 0
\end{bmatrix}
\begin{bmatrix} x(t) \\ x_a(t) \\ x_s(t) \end{bmatrix} \\ +
\begin{bmatrix} 0 & 0 \\ {I} & 0 \\ 0 & I
\end{bmatrix}
\begin{bmatrix} u(t) \\ u_s(t) \end{bmatrix} +
\begin{bmatrix}
w(t) \\ 0 \\ 0
\end{bmatrix} \\
y(t) = x_s(t)
\end{gathered}
\end{equation}
where $x_a$ and $x_s$ are virtual internal states corresponding to delayed actuator commands and delayed sensor signals, respectively, and $u_s$ represents compensation on virtual internal states. We can use standard control theory to obtain the optimal controller gain $K$ and optimal estimator gain $L$. Due to the block-matrix structure of the system matrices, the optimal gains have the following structure: $K = \begin{bmatrix}K_1 & K_2 & 0 \end{bmatrix}$, and $L = \begin{bmatrix}L_1^\top & 0 & L_2^\top \end{bmatrix}^\top$ \cite{Paper2}. The controller can be implemented as follows: 

\begin{equation}
\begin{gathered}
\delta(t+1) = Cx(t) - C\hat{x}(t) - L_2\delta(t) \\
\hat{x}(t+1) = A\hat{x}(t) + Bx_a(t) + L_1\delta(t) \\
u(t) = K_1\hat{x}(t) + K_2x_a(t)
\end{gathered}
\end{equation}

Here, $\delta$ is the delayed difference between the estimated sensor input and true sensor input, discounted by the observer term $L_2\delta(t)$. The resulting controller, shown in Fig. \ref{fig:of_delay_ifp}, contains two internal feedback pathways related to delay; one pathway compensates for sensor delays, and the other compensates for actuator delays. The remaining internal feedback is inherent to the Kalman Filter, as described in the previous section and shown in Fig. \ref{fig:of_standard_ifp}. Overall, the inclusion of sensor delays, actuator delays, and imperfect sensing result in an optimal controller with several internal feedback pathways, each of which serves a specific, interpretable purpose.

\subsection*{Localization of function requires lateral internal feedback}

Neuron-to-neuron computation and communication are typically spatially and temporally constrained for signals between areas, compared with signals within a local area. Although lateral feedback within each area may conform to the single-loop model, lateral feedback between areas are unexplained by the single-loop model, and are needed to to achieve good performance.

Localization of communication naturally leads to the localization of function in cortex; sensory and motor areas are spatially localized (with some cross-talk), as are different motor areas. 
We start with motor areas and the problem formulation described by  \eqref{eq:standard_system}, and partition tracking errors into two sets $x_1$ and $x_2$, representing two distinct but possibly coupled subsystems (e.g. two distinct limbs that are mechanically coupled). The overall tracking error is $x=\begin{bmatrix} x_1^\top & x_2^\top \end{bmatrix}^\top$. Correspondingly, we partition actuators into two sets $u_1$ and $u_2$ that act on their respective subsystems, via local controllers; $u=\begin{bmatrix} u_1^\top & u_2^\top \end{bmatrix}^\top$.

Each local controller senses and controls one subsystem; i.e. local controller 1 senses $x_1$ and computes $u_1$, and local controller 2 senses $x_2$ and computes $u_2$. Local controllers may communicate to one another; however, due to localization constraints, the cross-communication is delayed. Thus, local controller 1 cannot directly access $x_2$ without some delay, and similarly for local controller 2.

We first remark that without the constraint of localized communication, the optimal controller for Eq. \eqref{eq:standard_system} is $u=-Ax$. If $A$ is block-diagonal (i.e. $x_1$ and $x_2$ are uncoupled), then this controller obeys localized communication constraints; in fact, no cross-communication (internal feedback) is required between the two local controllers. However, if the two subsystems are coupled with time delays, then this controller requires rapid, global communication, which violate localized communication constraints. To enforce localized communication, we reformulate the problem by introducing virtual states $x_1^\prime$ and $x_2^\prime$, which represent delayed cross-communication between the two local controllers. $x_1^\prime$ is information sent from local controller 1 to local controller 2, with delay; and similarly for $x_2^\prime$. We also define $u_1^\prime$ and $u_2^\prime$, which model interconnections between virtual states and real tracking errors. For simplicity, we assume unit delay. The reformulated problem is as follows:

\begin{equation}
\begin{gathered}
\tilde{x} = \begin{bmatrix}x_1 \\ x_2^\prime \\x_1^\prime \\x_2 \end{bmatrix},
\tilde{u} = \begin{bmatrix}u_1 \\u_2^\prime \\u_1^\prime \\u_2 \end{bmatrix}
\tilde{w} = \begin{bmatrix}w_1 \\ 0 \\ 0 \\ w_2 \end{bmatrix}, \\
\tilde{A} = \begin{bmatrix}A_{11} & 0 & 0 & A_{12}\\0 & 0 & 0 & 0\\0 & 0 & 0 & 0\\A_{21} & 0 & 0 & A_{22}\end{bmatrix},
K = \begin{bmatrix}
\ast & \triangle & \ast & 0 \\
\triangle & \ast & \triangle & \ast \\
\ast & \triangle & \ast & \triangle\\
0 & \ast & \triangle & \ast
\end{bmatrix} \\
\tilde{x}(t+1) = \tilde{A}\tilde{x}(t) + \tilde{u}(t) + \tilde{w}(t) \\
\tilde{u} = K \tilde{x}
\end{gathered}
\end{equation}
The zeros in the top right and bottom left corners of the $K$ matrix preserve localized communication; they enforce that the two local controllers cannot communicate instantaneously to one another. Asterisks and triangles indicate free values; triangles represent sites of potential cross communication, or lateral internal feedback. We optimize over these free values to determine the $K$ matrix that achieves optimal performance with localized communication:

\begin{equation}K = \begin{bmatrix}-A_{11} & 0 & -A_{12} & 0 
\\0 & 0 & -A_{12} & A_{12}
\\A_{21} & -A_{21} & 0 & 0
\\0 & -A_{21} & 0 & -A_{22}\end{bmatrix}
\end{equation}

\begin{figure}
    \centering
    \includegraphics[page=6,width=0.8\linewidth]{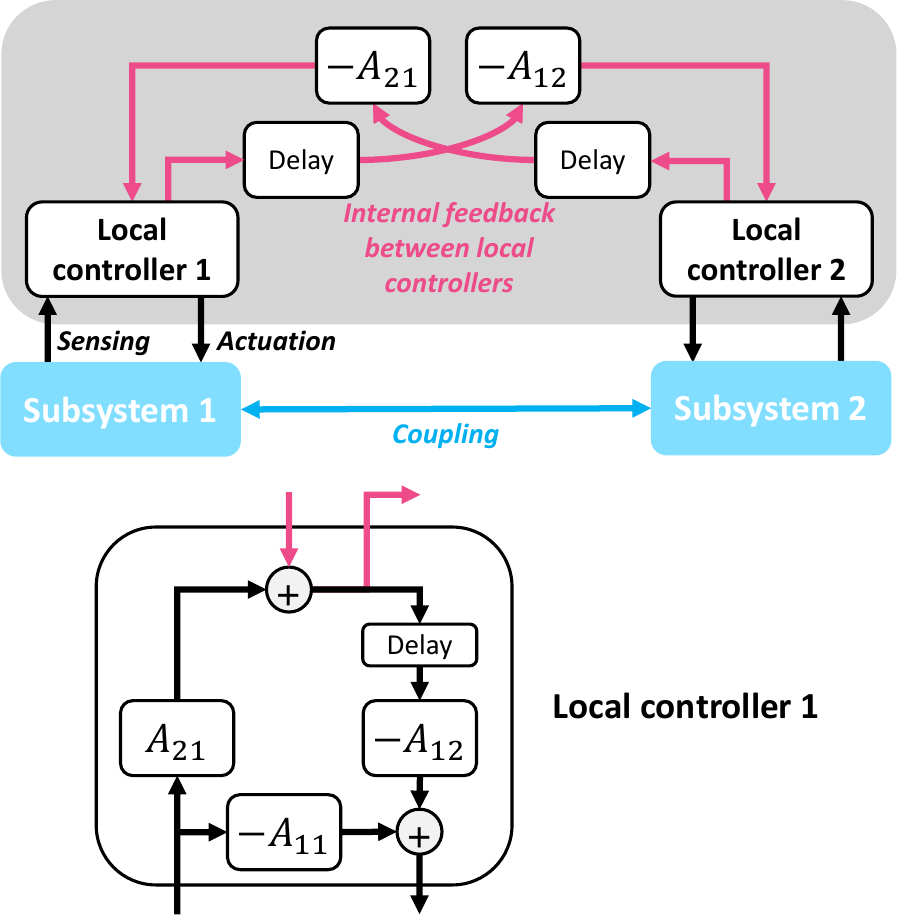}
    \caption{Optimal localized control of two coupled subsystems. \textit{(Top)} Overall schematic. Each subsystem has its own corresponding local controller, which senses and actuates only its assigned subsystem. Local controllers communicate to each other via lateral internal feedback (pink), with some delay. \textit{(Bottom)} Circuitry of local controller 1. Local controller 2 has identical circuitry, with different matrices; $A_{12}$ instead of $A_{21}$, $A_{22}$ instead of $A_{11}$, etc.}
    \label{fig:locality}
\end{figure}

The resulting local controllers are shown in Fig. \ref{fig:locality}. Note that the $-A_{12}$ term in the second row and the $-A_{21}$ term in the fourth row of $K$ correspond to lateral internal feedback. Here, these internal feedback signals carry predicted values of the unsensed tracking errors for each controller, after taking control action into account; for instance, internal feedback from local controller 2 to local controller 1 conveys the predicted value of $x_2$, after taking control action from controller 2 into account. We can develop intuition for this implementation by following an impulse $w$ through time:
\begin{equation}   
\begin{gathered}
\tilde{x}(1) = \begin{bmatrix}w_1 \\ 0 \\ 0 \\ w_2\end{bmatrix}\rightarrow \tilde{x}(2) = \begin{bmatrix}A_{12}w_2 \\ A_{12}w_2 \\ A_{21}w_1 \\ A_{21}w_1\end{bmatrix} \rightarrow \tilde{x}(3) = 0
\end{gathered}
\end{equation}

\begin{figure}
    \centering
    \includegraphics[page=9,width=0.7\linewidth]{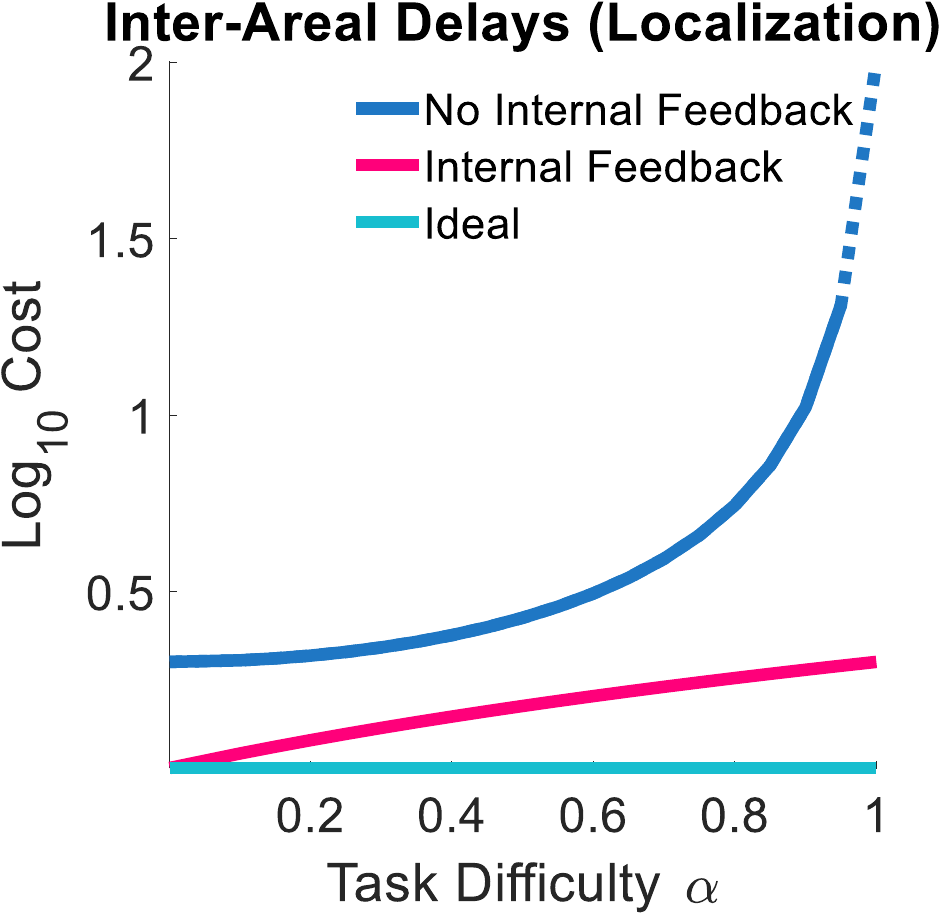}
	\caption{Localization of function within motor-related cortex: although different parts of the cortex control different parts of the body, these parts of the body are inherently mechanically coupled. As a result, internal feedback is useful and in some cases necessary to maintain localization of function. In simulations, we consider the problem of tracking a moving target over a two-dimensional space, varying the task difficulty. The 'Ideal' controller is centralized (i.e. no delays between local controllers) and obtains the best performance. The localized controller with internal feedback achieves similar performance. The localized controller without internal feedback suffers from substantially worse performance (higher cost). As task difficulty increases, the task becomes infeasible without internal feedback (broken line). \label{fig:neurolocalization}} 
\end{figure}

We compare the performance of this controller to the controller without internal feedback in Fig. \ref{fig:neurolocalization}. For the controller without internal feedback, we choose the best possible linear controller; however, the lack of internal feedback results in severe performance degradation. As task difficulty increases, this controller is unable to stabilize the closed-loop system and tracking becomes infeasible. With internal feedback, task performance stays near the centralized optimal (i.e. the case where local controllers can communicate freely without delay).  

The foregoing analysis shows that localization of motor function (i.e. specialization of parts of motor cortex to particular parts of the body) in fact requires cross-communication, or internal feedback, between local controllers. Local circuits in the two hemispheres must also be coordinated -- indeed, they are connected by a massive corpus collosum that crosses the midline.  The cross talk between local controllers is supported by the presence of signals relating to the whole body in parts of the motor cortex specialized to particular parts of the body. Internal feedback enables localization of function when subsystems are coupled; in reality, all body movements are mechanically coupled, something which the motor system can conceal through effective localization and coordination.

\subsection*{Speed-accuracy trade-offs necessitate the use of layering and internal feedback for attention}

We have shown that state estimation and localization of function require internal feedback to correct for self-generated or predictable movements. We now consider the role of \textit{attention} in the context of a tracking task. Our model can be considered an implementation by internal feedback of the observation that attention enhances the responses of neurons that selectively respond to an attended stimulus \cite{reynolds2000}.

Up to this point, we have assumed that the controller can directly sense the position of the object (perhaps with some delay). In a real world a scene can have many objects, which makes it more difficult for a sensorimotor system to determine the position of an object in the scene. However, a moving object, once identified, can be discriminated from a static visual scene. This illustrates the distinction between scene-related tasks (such as object identification) and error-related tasks (such as object tracking), which in the visual cortex is accomplished by the ventral and dorsal streams, respectively. 

This distinction also mirrors the separation between bumps and trails in the mountain-biking task studied in \cite{Nakahira2021}, allowing us to build on the control architecture in that task. The main difference is that instead of separating into two control loops, we use layering and internal feedback to supplement the control actions of the main control loop. 

For simplicity of presentation, we consider a one-dimensional problem (tracking on a line), and use as the metric $\|x\|_\infty$ (worst-case, or adversarial tracking error) rather than $\|x\|_2$ (average-case tracking error). We have some object whose position relative to us, $r$, is governed by the dynamics 
\begin{equation}
r(t+1) = r(t) + w_r(t) + w_b(t)
\end{equation}
where $w_r$ represents object movement, and $w_b$ represents changes in the background scene. Our limb position $p$ is governed by the dynamics
\begin{equation}
p(t+1) = p(t) - u(t)
\end{equation}
where $u(t)$ is some limb action. The tracking error $x := r - p$ then obeys the dynamics
\begin{equation}
x(t+1) = x(t) + w_r(t) + w_b(t) + u(t)
\end{equation}
where the task difficulty is implicitly equal to 1.

We assume that object movement and background changes are bounded: $|w_r(t)| \leq \epsilon_r$ and $|w_b(t)| \leq \epsilon_b$ for all $t$. Additionally, we assume that background changes are much slower than object movement: $\epsilon_b \ll \epsilon_r$, i.e. 
\begin{equation} \label{eq:eps_approx}
\epsilon_r + \epsilon_b \approx \epsilon_r
\end{equation}

Consider a movable sensor that senses some interval of size $\beta$ on the continuous line. Information from the sensor must be communicated to the controller (via axons), but this communication is subject to speed-accuracy trade-offs. Let $R$ be the signaling rate (bits per unit time), let $T$ be the signaling delay, and let $\lambda$ be the resource cost to build and maintain axons. The speed-accuracy trade-off can be formalized as $R = \lambda T$ \cite{Nakahira2015}. The signaling rate $R$ is related to the resolution of information sent about the sensed interval. 

\begin{figure}
    \centering
    \includegraphics[page=7,width=0.8\linewidth]{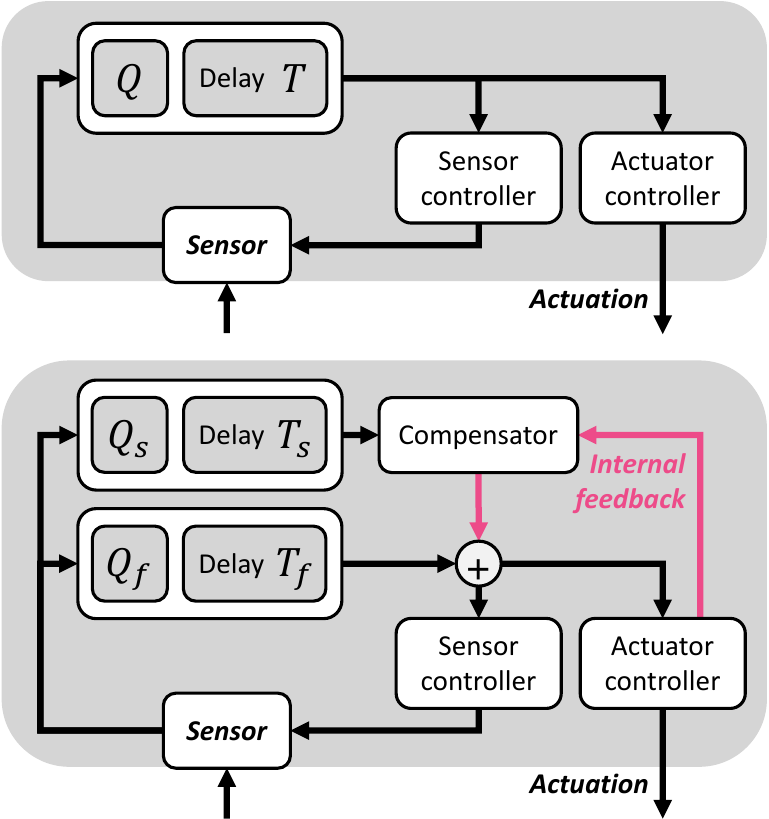}
    \caption{Optimal control model of attention, with moveable sensor. \textit{(Top)} Model with one communication path, in which information is quantized by quantizer $Q$ and conveyed to the controller with delay $T$. \textit{(Bottom)} Model with two communication paths, and two separate quantizers $Q_s$, $Q_f$ and respective delays. This model uses lateral internal feedback (pink) between the two controller paths}
    \label{fig:attention_diagram}
\end{figure}

We implement this speed-accuracy trade-off using a static, memoryless quantizer $Q$ with uniform partition, followed by a communication delay, as shown on the top in Fig. \ref{fig:attention_diagram}. This choice of quantizer does not add to the cost, since it recovers the optimal cost over all possible quantizers \cite{sarma_flexibility_2019}. The controller can move the sensor around; the interval sensed by the sensor remains constant, but the controller can choose where the interval lies. Assume the initial position of the object is known -- we can select an initial sensor location and $\beta$ appropriately such that $r(t)$ always falls within the sensed interval. In this case, the best possible tracking error for any delay $T$ is
\begin{equation} \label{eq:one_layer}
\epsilon_r T + \frac{\beta}{2^{\lambda T}}
\end{equation}
The first term represents error from delay, object movement, and drift. In the time taken for information to reach the controller, the most adversarial action possible by the object and background would contribute a tracking error of $(\epsilon_r + \epsilon_b)T$; we apply simplification \eqref{eq:eps_approx} to obtain $\epsilon_r T$. The second term represents quantization error. For an interval of size $\beta$ divided into $N$ uniform sub-intervals, the worst-case error is $\frac{\beta}{N}$; we then use the fact that $N=2^R=2^{\lambda T}$.

\begin{figure}
\centering
\includegraphics[page=11,width=\linewidth]{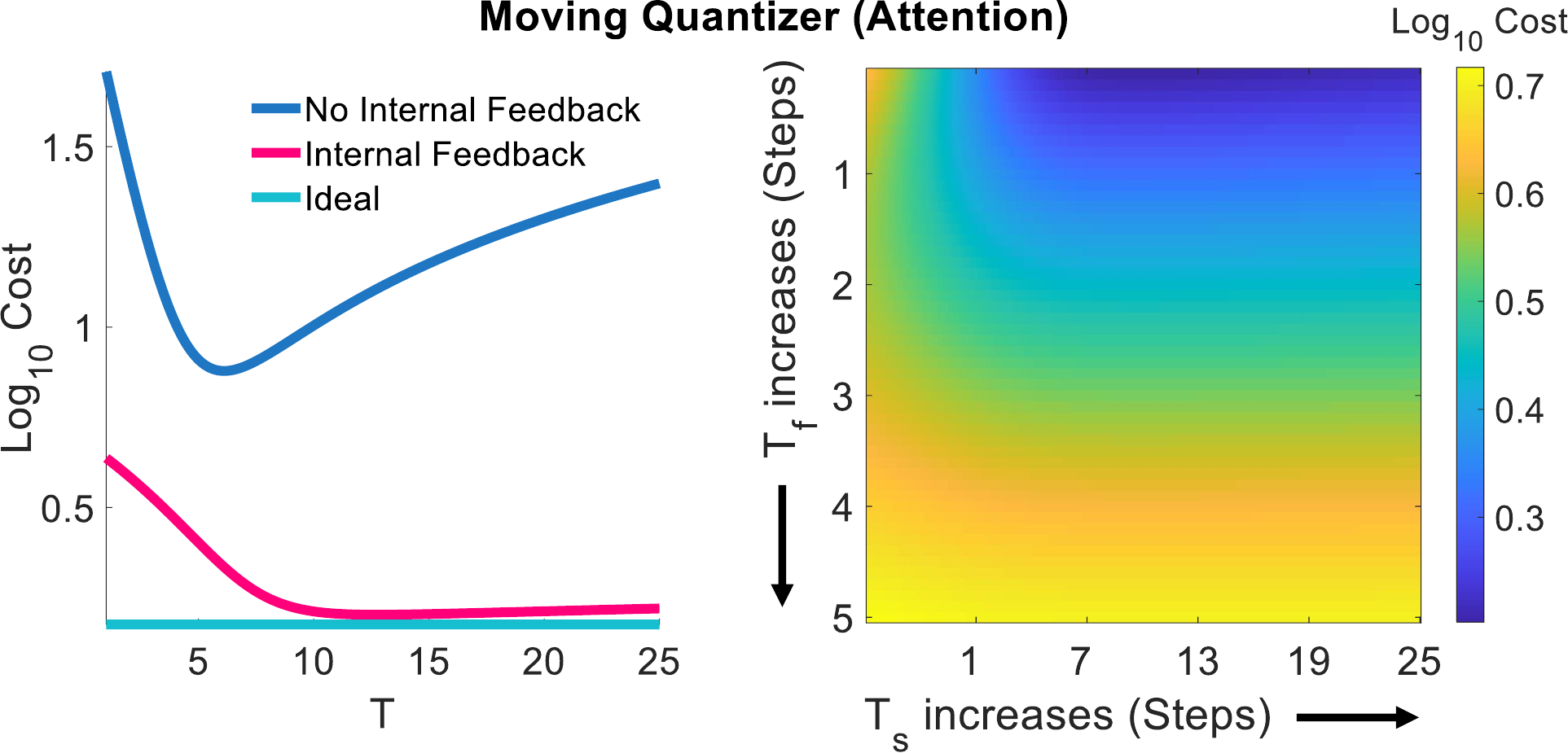}
\caption{\textit{(Left)} Internal feedback and layering achieve superior performance when sensor-controller communications are subject to speed-accuracy trade-offs. The `No Internal Feedback' controller uses one layer, while the `Internal Feedback' controller uses two layers, with internal feedback between the layers. The two-layer case consists of a fast forward pathway compensated by slow internal feedback, which takes slow background changes into account; this achieves better performance (lower cost) than the case without internal feedback. The `Ideal' controller, where the sensor directly senses the moving object, is also shown. The layered system with internal feedback achieves performance close to the ideal. Task difficulty is $\alpha = 1$. $T$ represents delay. For the `No Internal Feedback' controller, it represents the delay of the single layer; for the `Internal Feedback' controller, it represents the delay of the slow layer, i.e. $T = T_s$. The delay of the fast layer is held constant. \textit{(Right)} Performance (log cost) of the two-layer controller with internal feedback as delays of both layers are varied. Notice that performance is generally good when $T_f$ is low and $T_s$ is sufficiently high.}\label{fig:neuroattention}
\end{figure}

This is achieved by the controller depicted on the top in Fig. \ref{fig:attention_diagram}. The cost, as a function of $T$, is plotted in the left panel Fig. \ref{fig:neuroattention} with the label `No Internal Feedback' (where $T=T_s$). Here, the speed-accuracy trade-off is implicit. Very low values of $T$ correspond to very low signaling rates -- the controller does not receive enough information to act accurately, so performance is poor. The opposite problem occurs at very high values of $T$; though the information is high-resolution, the time elapsed between information and action is too long, leading to poor performance. The best performance occurs between these two extremes.

We can improve this performance by nearly an order of magnitude by adding an additional communication pathway and the requisite internal feedback. We now have two communication paths from the sensor, each with its own quantizer and delay block. The slower communication path uses quantizer $Q_s$ with delay $T_s$, while the faster path uses $Q_f$ with delay $T_f$. To further facilitate speed in the fast path, we allow it to send only a subset of information from the sensor (i.e. only send information about a small part of the sensed scene). Mathematically, let the fast path send information about an interval of size $\beta_f$, with $\beta_f < \beta$, and let this smaller sub-interval be contained within the sensor interval. This sub-interval is an implementation of \textit{attention}. The fast path is the main actuation path, while the slower path provides compensatory signals via internal feedback; this is shown on the bottom in Fig. \ref{fig:attention_diagram}. In this case, the best possible cost is
\begin{equation}
\begin{gathered}
\epsilon_r T_f + \frac{\beta_f + E_s}{2^{\lambda T_f}} \\
E_s = \epsilon_b T_s + \frac{\beta}{2^{\lambda T_s}}
\end{gathered}
\end{equation}
The first term represents error from delay and object movement, similar to \eqref{eq:one_layer}. The second term represents a combination of quantization error from the fast communication pathway ($\beta_f$) and performance error of the slow pathway ($E_s$), which informs the fast pathway of where to place the sub-interval. Notice that $E_s$ takes the same form as \eqref{eq:one_layer}. 

The cost, as a function of $T_s$, is plotted in the left panel of Fig. \ref{fig:neuroattention} with the label `Internal Feedback'. In this plot, we assume $T_f$ to be its smallest possible value; one unit of delay. We see that using two quantizers in combination with internal feedback is superior to using one quantizer. We remark that this only holds when the two quantizers are \textit{different}; if we simply use two quantizers with the same interval, bit rate, and delay, no such performance boost occurs. In general, holding $T_s$ constant and decreasing $T_f$ improves performance, as shown in the right panel of Fig. \ref{fig:neuroattention}. 

Functionally, the inclusion of a faster communication pathway allows action to be taken in a more timely manner than in the single-pathway case. Unlike in the single-pathway case, we are not encumbered by issues of low-resolution information; the slower communication pathway corrects the fast pathway through internal feedback. Here, as in previous examples, the internal feedback carries signals correcting for self-generated and slow, predictable changes. Overall, despite speed-accuracy trade-offs in communication, the system achieves fast and accurate behavior with the help of internal feedback, under reasonable assumptions about the dynamics of the scene and environment.  

\section*{Discussion}

We analyzed a group of basic control models to explore internal feedback in an action-perception loop with time delays and limited communication bandwidth. What emerged was mathematical explanations for previously cryptic features of biological sensorimotor control. We showed how internal feedback is required for state estimation and localization of function, and how it facilitates attention and improved motor performance. This is a first step toward an end-to-end model of sensorimotor processing. 



\subsection*{Existing frameworks for internal feedback neglect task performance}
There may be other functions for internal feedback in addition to compensating for time delays and limited cortical communication bandwidth. Additional functions that have been suggested are computation through dynamics \citep{VyasShenoy2020}, deploying recurrent networks \citep{KarDiCarlo2019}, performing Bayesian inference \citep{BastosFriston2012,Swain07}, generating predictive codes \citep{RaoBallard1999,Keller2018}, and many others. These frameworks have arisen in parallel with the development of methods for increasingly high-throughput and high-resolution measurements of biological systems, which support the idea that internal feedback and internal dynamics are essential features of cortical computation. 

These frameworks emphasize prediction but are largely confined to the context of either sensory processing or motor processing separately and do not explicitly model ethological task performance. Our analysis emphasizes motor-based internal feedback signals, which have not been considered in previous frameworks. Recent data showing that motor signals and effects of past and current actions account for substantial cortical activity, previously considered spontaneous, background or noise, are consistent with this view \cite{Willett2020, Stavisky2019, Stringer2019, Musall2019,Leinweber2017}. The ultimate purpose of sensory processing is to support optimal actions for ensuring survival \cite{Churchland1994}. 

We propose that internal feedback signals carry information about how actions propagate through the body and its environment, as well as about planned future actions, including how communication limitations affect both plans and actions. Our implementation of control produces performance that is nearly Bayes optimal, illustrating the difference between behavior and implementation introduced earlier. Our emphasis on this motor-centric view of the cortex complements the sensory-centric view that has been explored much more thoroughly to date.

\subsection*{Existing optimal control models neglect physiological limitations}  
Optimal control theory is a general framework for sensorimotor modeling. Given a mathematical description of a system and some task specification, the optimal controller provably gives the best possible performance. However, these proofs assume that the components are fast and accurate, with communication at the speed of light and control circuits implemented with fast and accurate electronics; with these components, a single sense-compute-actuate loop is generally sufficient to achieve optimal behavior. 

To use control theory to model physiological circuitry, a distinction between \textit{behavior} and \textit{implementation} must be made. The same behavior (optimal performance) may be achieved through a number of different implementations (underlying circuitry). Although traditional control theory excels as a model of sensorimotor behavior, it does not incorporate the component-level constraints that are prevalent in biology; as a consequence, the implementations of traditional control theory models are not directly relevant to biological control. 

Recent advances have extended traditional control theory to allow distributed control and incorporation of component-level constraints \citep{Nakahira2021,Anderson2019,Paper2,Paper3}. We build on this body of work to describe how constrained components affect the implementation of an optimal distributed biological controller. In particular, we show that internal feedback is a necessary part of any controller whose components exhibit the speed-accuracy trade-offs found in brains. 

Fast long-range association fibers in the cortex are metabolically and developmentally expensive, have low bandwidth (compared to counterfactual slower fibers), require constant maintenance and repair and are limited in number. Internal feedback from the motor cortex to earlier sensory areas can regulate the use of these pathways by suppressing self-generated and other predictable signals and using the fast pathways to selectively transmit the unpredicted changes needed by the motor system to make make fast decisions. This virtualizes the behavior of the control system to produce actions that are both fast and accurate despite the internal time delays and limited communication bandwidth.   

\subsection*{Biophysical speed-accuracy trade-offs necessitate internal feedback}
Biological control systems do not have components that are both fast and accurate. Spiking neurons, though fast relative to other biological signaling mechanisms, are many orders of magnitude slower than electronics and face severe speed-accuracy trade-offs that constrain communication and control. However, by cleverly combining components with different speed-accuracy trade-offs and using internal feedback as demonstrated above, brains are able to perform survival-critical sensorimotor tasks with speed and accuracy.

Neurons have biophysical constraints that lead to speed-accuracy trade-offs; for example, some neurons can rapidly convey a few bits of information, and others can slowly convey many bits of information, but neurons that rapidly convey many bits of information are expensive and correspondingly rare. Speed-accuracy trade-offs include spike averaging versus spike timing and the spatial trade-off between the number of neurons (information rate) and their axonal diameter (conduction speed) in nerve bundles \cite{Sterling2015, Nakahira2021}. These trade-offs have consequences for the performance of sensorimotor systems that we can study in our control models.

Brains contain highly diverse populations of neurons; for example, the range of neural conduction speeds in humans spans several orders of magnitude \cite{Sterling2015}. In sensorimotor systems, these diverse neurons are multiplexed in a task-specific way that approximates the performance of a single-loop system composed of ideal (e.g. fast \textit{and} accurate) components \cite{Nakahira2021}. The fastest components are used in the feedforward loop, sending information from sensing areas toward motor areas. Internal feedback compensates for accuracy by filtering out slow-changing, predictable, or task-irrelevant stimuli, such that the fewest possible bits need to travel along the fastest possible neurons. From an evolutionary perspective, once a system can achieve fast responses, additional layers of control can be added to achieve more accurate and flexible behavior without sacrificing performance.

The reason why internal feedback is not needed in most engineered control systems is that internal time delays are negligible. But in biological systems, even the fastest neurons used in the feedforward loop give rise to significant time delays. This is why it is essential to include delays in our control-based analyses of the forward loop (Fig. \ref{fig:fc_ifp}).

\subsection*{Fast forward conduction is key to successful sensorimotor task performance}
In cortex, the fastest, largest and most striking neurons are the large pyramidal cells: Meynert cells in primary visual cortex carry signals from rapidly moving-object; giant Betz cells in motor cortex that project to the spinal cord are responsible for rapid responses to perturbations from planned movements; and giant Von Economo cells in the prefrontal cortex (anterior cingulate and fronto-insular areas) that project subcortically are involved in the regulation of emotional and cognitive behavior.  \cite{Livingstone_1998,ChanPalay1974,Fetz1984,allman2010}. 

The visual stream diverges into the dorsal and ventral streams, which are responsible for object motion and object identity, respectively. In natural scenes, object locations may change quickly, but object identities are change relatively slowly; a mouse may move around rapidly, but its status as a prey hunted by a barn owl does not change. Thus, speed is crucial for the dorsal stream, but not the ventral stream. 

This difference has physiological consequences in our minimal model of attention that could explain differences between cortical projections in the two streams: the giant Meynert Cells that project from V1 to MT (an object motion area in the dorsal stream; see Fig. \ref{fig:single_loop}), but there are no equivalently large cells projecting from V1 to inferotemporal cortex (an object identity area in the ventral stream). 
Reaching tasks could test the predictions of our control model for rapidly and unpredictably moving objects on a fixed background compared with predictably moving objects on nonstationary backgrounds.

Neurons in MT respond selectively to the direction of moving objects and provide signals that are used by the oculomotor system for the smooth pursuit of moving objects \cite{lisberger2010}.There are several visual pathways from the retina to the cortex for tracking moving stimuli.  In addition to the cortical pathway that projects from V1 to area MT in Fig. \ref{fig:single_loop}, the retina also projects to the pulvinar, another thalamic relay to extrastriate areas of the cortex \cite{warner2010}.  These two pathways could implement the optical control model in the bottom panel of Fig. \ref{fig:attention_diagram}, where the fast, direct pathway is from the pulvinar and the delayed, indirect pathway from V1 corresponds to the slower pathway.

\subsection*{Internal feedback facilitates fast feedforward signals in visual cortex}
In recent years, large-scale recordings from visual cortex have uncovered non-visual signals that challenge the traditional single-loop view of sensorimotor control. In the traditional view, 
visuomotor processing consists of a series of successive transformations from stimulus to response, with each cortical area along the way tuned to some aspect of stimulus space \cite{Hubel_Wiesel_1959}. However, although V1 does respond to visual stimuli, the activity of these neurons is dominated by motor-based internal feedback and task/attention-related modulatory internal feedback \cite{Churchland1994, Stringer2019, Musall2019, Muckli2013}. 

The number of projections from V1 to V2 is roughly the same as number of neurons, of similar conduction speed, that project from V2 to V1 \cite{Callaway2004,Angelucci2003,ElShamaylehKumbhani2013}. However, these neurons are very different in morphological and molecular characteristics: The neurons that project feedforward from V1 to V2 primarily activate AMPA receptors, while the feedback neurons that project from V2 to V1 have a strong NMDA receptor component and terminate almost exclusively on excitatory pyramidal neurons \cite{Self2012, anderson2016}. Both of these receptors are activated by glutamate, but AMPA-mediated currents are fast, lasting only a few ms, while NMDA-mediated currents can linger in the postsynaptic neurons for hundreds of ms  \cite{Attwell2005}. This feedback could be relevant for top-down signaling to shape and control perception during actions and could also be important for perceptual learning. 

Pharmacologically blocking NMDA receptors in visual cortex disrupts figure-ground discrimination; that is, a loss in capacity to contextually interpret the visual scene \cite{widmer2022}.  In the context of our theory and minimal model of attention, internal feedback from V2 informs V1  of predictable elements in future stimuli. 
Since the visual space cannot be sampled losslessly, these feedback signals could be helping V1 suppress predictable features, making the the unpredictable features more salient \cite{Self2012}.

\subsection*{Internal feedback facilitates localization of function in motor cortex}
Primary motor cortex (M1) 
is dominated by its own past activity rather than static representations \cite{Churchland2012}. In the context of the state estimation problem we considered in Figure \ref{fig:of_delay_ifp}, motor cortex is dominated neither by motor representations nor by pattern generation, but by predictions of the consequences of self-action through local internal feedback, which need to be sent throughout the body. 

By the same principle, the localization of function within motor cortex that we considered in Figure \ref{fig:neurolocalization} reconciles  the conventional view of homuncular organization with, for example, the  body-related signals found in putatively hand-related parts of motor cortex, as well as contralateral hand signals \citep{Stavisky2019, Willett2020, Ames2019}. As with motor signals in visual cortex, these broad body movement signals in motor cortex 
are \textit{necessary} for identifying predictable consequences of motor signals from other body movements and separating them from unpredictable signals of critical importance for rapidly controlling localized body parts. This provides each body part with the context it needs to compensate for the movement of other body parts.

\subsection*{Learning on internal feedback pathways fine-tunes performance} 
Internal feedback pathways carry attentional signals that activate slow NMDA receptors, which in turn regulate the strengths of synapses \cite{li2009}. 
We have shown that internal feedback pathways are needed for ignoring self-generated and other predictable signals.
Early in brain development, activation of NMDA receptors in primary visual cortex before the first visuomotor experience is needed to  suppress predictable feedback and the the selection of unpredictable stimuli \cite{widmer2022}. Knocking out these NMDA receptors impairs visuomotor skill learning later in life, necessary for compensating for body growth and body weight changes. This form of learning is driven by prediction error. Thus, the same internal feedback system that broadcasts predictions for upcoming actions could drive learning through local prediction error.

Reinforcement learning governed by circuits in the basal ganglia may also benefit from the internal feedback pathways in the cortex. Reinforcement is a scalar signal that does not specify which sensory inputs were responsible for the reward, which in part is why it is a much slower form of learning. Attentional internal feedback in the cortex automatically selects and represents the currently most salient information for motor actions. This information projects to the striatum and makes it easier for the basal ganglia to associate the causally relevant sensory inputs with reward signals \cite{Churchland1994}.

Attention has been studied primarily in the context of sensory processing.  The importance of attentional signals for reducing time delays in making motor decisions adds a new direction for future experimental studies. Attention is linked to conscious awareness and rides atop the global representation of the body throughout the cortex. This makes internal feedback a candidate feature of the nervous system that helps explain  the sense of unity that we experience, which would otherwise be difficult to achieve with a balkanized representation of body parts.




\acknow{Stephen Lisberger helped clarify our discussion on visual pathways to the oculomotor system. J.C.D. and T.J.S. were supported by NSF NCS-FO 1735004. T.J.S. was supported by ONR N00014-16-1-282. J.S.L. was in part supported by NSERC PGSD3-557385-2021. This paper is based on the doctoral research of A.A.S. and J.S.L.}

\showacknow{} 

\bibliography{main}

\begin{thebibliography}{10}

\bibitem{doyle_feedback_1992}
J Doyle, B Francis, A Tannenbaum, {\em Feedback {Control} {Theory}}.
\newblock (Macmillan), (1992).

\bibitem{Wolpert1995}
DM Wolpert, Z Ghahramani, MI Jordan, An internal model for sensorimotor
  integration.
\newblock {\em\protect\JournalTitle{Science}} \textbf{269}, 1880–1882 (1995).

\bibitem{Todorov2004}
E Todorov, {Optimality principles in sensorimotor control}.
\newblock {\em\protect\JournalTitle{Nature Neuroscience}} \textbf{7}, 907--915
  (2004).

\bibitem{Franklin2011}
DW Franklin, DM Wolpert, Computational mechanisms of sensorimotor control.
\newblock {\em\protect\JournalTitle{Neuron}} \textbf{72}, 425--442 (2011).

\bibitem{zagha2020}
E Zagha, Shaping the cortical landscape: Functions and mechanisms of top-down
  cortical feedback pathways.
\newblock {\em\protect\JournalTitle{Front Syst Neurosci.}} \textbf{14}, 33
  (2020).

\bibitem{Zhaoping2014}
L Zhaoping, {\em Understanding Vision}.
\newblock (Oxford University Press), (2014).

\bibitem{GollischMeister2010}
T Gollisch, M Meister, Eye smarter than scientists believed: Neural
  computations in circuits of the retina.
\newblock {\em\protect\JournalTitle{Neuron}} \textbf{65}, 150–164 (2010).

\bibitem{Callaway2004}
EM Callaway, {Feedforward, feedback and inhibitory connections in primate
  visual cortex}.
\newblock {\em\protect\JournalTitle{Neural Networks}} \textbf{17}, 625--632
  (2004).

\bibitem{Angelucci2003}
A Angelucci, J Bullier, {Reaching beyond the classical receptive field of V1
  neurons: Horizontal or feedback axons?}
\newblock {\em\protect\JournalTitle{Journal of Physiology Paris}} \textbf{97},
  141--154 (2003).

\bibitem{ElShamaylehKumbhani2013}
Y El-Shamayleh, RD Kumbhani, NT Dhruv, JA Movshon, Visual response properties
  of v1 neurons projecting to v2 in macaque.
\newblock {\em\protect\JournalTitle{Journal of Neuroscience}} \textbf{33},
  16594–16605 (2013).

\bibitem{Felleman1991}
DJ Felleman, DC {Van Essen}, {Distributed hierarchical processing in the
  primate cerebral cortex}.
\newblock {\em\protect\JournalTitle{Cerebral Cortex}} \textbf{1}, 1--47 (1991).

\bibitem{Muckli2013}
L Muckli, LS Petro, {Network interactions: non-geniculate input to V1}.
\newblock {\em\protect\JournalTitle{Current Opinion in Neurobiology}}
  \textbf{23}, 195--201 (2013).

\bibitem{Churchland2012}
MM Churchland, et~al., Neural population dynamics during reaching.
\newblock {\em\protect\JournalTitle{Nature}} \textbf{487}, 51–56 (2012).

\bibitem{Willett2020}
FR Willett, et~al., Hand knob area of premotor cortex represents the whole body
  in a compositional way (2020).

\bibitem{Stavisky2019}
SD Stavisky, et~al., Neural ensemble dynamics in dorsal motor cortex during
  speech in people with paralysis (2019).

\bibitem{Stringer2019}
C Stringer, et~al., {Spontaneous behaviors drive multidimensional, brainwide
  activity}.
\newblock {\em\protect\JournalTitle{Science}} \textbf{364} (2019).

\bibitem{Musall2019}
S Musall, MT Kaufman, AL Juavinett, S Gluf, AK Churchland, {Single-trial neural
  dynamics are dominated by richly varied movements}.
\newblock {\em\protect\JournalTitle{Nature Neuroscience}} \textbf{22},
  1677--1686 (2019).

\bibitem{Ebrahimi2022}
S Ebrahimi, et~al., Emergent reliability in sensory cortical coding and
  inter-area communication.
\newblock {\em\protect\JournalTitle{Nature}} \textbf{605}, 713--721 (2022).

\bibitem{RaoBallard1999}
RPN Rao, DH Ballard, Predictive coding in the visual cortex: a functional
  interpretation of some extra-classical receptive-field effects.
\newblock {\em\protect\JournalTitle{Nature Neuroscience}} \textbf{2}, 79–87
  (1999).

\bibitem{Keller2018}
GB Keller, TD Mrsic-Flogel, Predictive processing: A canonical cortical
  computation (2018).

\bibitem{VyasShenoy2020}
S Vyas, MD Golub, D Sussillo, KV Shenoy, Computation through neural population
  dynamics.
\newblock {\em\protect\JournalTitle{Annual Review of Neuroscience}}
  \textbf{43}, 249–275 (2020).

\bibitem{KarDiCarlo2019}
K Kar, J Kubilius, K Schmidt, EB Issa, JJ DiCarlo, Evidence that recurrent
  circuits are critical to the ventral stream’s execution of core object
  recognition behavior.
\newblock {\em\protect\JournalTitle{Nature Neuroscience}} \textbf{22},
  974–983 (2019).

\bibitem{BastosFriston2012}
A Bastos, et~al., Canonical microcircuits for predictive coding.
\newblock {\em\protect\JournalTitle{Neuron}} \textbf{76}, 695–711 (2012).

\bibitem{Swain07}
E Libby, TJ Perkins, PS Swain, Noisy information processing through
  transcriptional regulation.
\newblock {\em\protect\JournalTitle{Proceedings of the National Academy of
  Sciences}} \textbf{104}, 7151–7156 (2007).

\bibitem{Anderson2019}
J Anderson, JC Doyle, SH Low, N Matni, System level synthesis.
\newblock {\em\protect\JournalTitle{Annual Reviews in Control}} \textbf{47},
  364--393 (2019).

\bibitem{Nakahira2021}
Y Nakahira, Q Liu, TJ Sejnowski, JC Doyle, Diversity-enabled sweet spots in
  layered architectures and speed–accuracy trade-offs in sensorimotor
  control.
\newblock {\em\protect\JournalTitle{Proceedings of the National Academy of
  Sciences of the United States of America}} \textbf{118}, 1--11 (2021).

\bibitem{Paper1}
AA Sarma, et~al., Internal feedback in biological control: Architectures and
  examples in {\em Proceedings of the IEEE American Control Conference}.
\newblock pp. 456--461 (2022).

\bibitem{Paper2}
J Stenberg, JS Li, AA Sarma, JC Doyle, Internal feedback in biological control:
  Diversity, delays, and standard theory in {\em Proceedings of the IEEE
  American Control Conference}.
\newblock pp. 462--467 (2022).

\bibitem{Paper3}
JS Li, Internal feedback in biological control: Locality and system level
  synthesis in {\em Proceedings of the IEEE American Control Conference}.
\newblock pp. 474--479 (2022).

\bibitem{reynolds2000}
J Reynolds, T Pasternak, R Desimone, Attention increases sensitivity of v4
  neurons.
\newblock {\em\protect\JournalTitle{Neuron}} \textbf{26}, 703--714 (2000).

\bibitem{Nakahira2015}
Y Nakahira, N Matni, JC Doyle, {Hard limits on robust control over delayed and
  quantized communication channels with applications to sensorimotor control}
  in {\em Proc. IEEE CDC}.
\newblock pp. 7522--7529 (2015).

\bibitem{sarma_flexibility_2019}
AA Sarma, JC Doyle, Flexibility and cost-sensitivity in a quantized control
  loop in {\em Proceedings of the IEEE American Control Conference}.
\newblock (2019).

\bibitem{Leinweber2017}
M Leinweber, DR Ward, JM Sobczak, A Attinger, GB Keller, A sensorimotor circuit
  in mouse cortex for visual flow predictions (2017).

\bibitem{Churchland1994}
PS Churchland, VS Ramachandran, TJ Sejnowski, A critique of pure vision in {\em
  Large-Scale Neuronal Theories of the Brain}, eds.{} C Koch, J Davis.
\newblock (MIT Press), pp. 23--60 (1994).

\bibitem{Sterling2015}
P Sterling, SB Laughlin, {\em {Principles of neural design}}.
\newblock (MIT Press), (2015).

\bibitem{Livingstone_1998}
MS Livingstone, Mechanisms of direction selectivity in macaque v1.
\newblock {\em\protect\JournalTitle{Neuron}} \textbf{20}, 509–526 (1998).

\bibitem{ChanPalay1974}
V Chan-Palay, SL Palay, SM Billings-Gagliardi, Meynert cells in the primate
  visual cortex.
\newblock {\em\protect\JournalTitle{Journal of Neurocytology}} \textbf{3},
  631–658 (1974).

\bibitem{Fetz1984}
EE Fetz, Functional organization of motor and sensory cortex: symmetries and
  parallels in {\em Dynamic Aspects Of Neocortical Function}, ed.{} WC
  G.M.~Edelman, W.E.~Gall.
\newblock (John Wiley), pp. 453--474 (1984).

\bibitem{allman2010}
J Allman, et~al., The von economo neurons in frontoinsular and anterior
  cingulate cortex in great apes and humans.
\newblock {\em\protect\JournalTitle{Brain Structure and Function}}
  \textbf{214}, 495--517 (2010).

\bibitem{lisberger2010}
SG Lisberger, Visual guidance of smooth-pursuit eye movements: sensation,
  action, and what happens in between.
\newblock {\em\protect\JournalTitle{Neuron}} \textbf{66}, 477--491 (2010).

\bibitem{warner2010}
CE Warner, Y Goldshmit, JA Bourne, Retinal afferents synapse with relay cells
  targeting the middle temporal area in the pulvinar and lateral geniculate
  nuclei.
\newblock {\em\protect\JournalTitle{Front Neuroanat}} \textbf{4}, 8 (2010).

\bibitem{Hubel_Wiesel_1959}
DH Hubel, TN Wiesel, Receptive fields of single neurones in the cat’s striate
  cortex (1959).

\bibitem{Self2012}
MW Self, RN Kooijmans, H Sup{\`{e}}r, VA Lamme, PR Roelfsema, {Different
  glutamate receptors convey feedforward and recurrent processing in macaque
  V1}.
\newblock {\em\protect\JournalTitle{Proceedings of the National Academy of
  Sciences of the United States of America}} \textbf{109}, 11031--11036 (2012).

\bibitem{anderson2016}
J Anderson, K Martin, Chapter 6 - interareal connections of the macaque cortex:
  How neocortex talks to itself in {\em Axons and Brain Architecture}, ed.{} KS
  Rockland.
\newblock (Academic Press, San Diego), pp. 117--134 (2016).

\bibitem{Attwell2005}
D Attwell, A Gibb, {Neuroenergetics and the kinetic design of excitatory
  synapses}.
\newblock {\em\protect\JournalTitle{Nature Reviews Neuroscience}} \textbf{6},
  841--849 (2005).

\bibitem{widmer2022}
FC Widmer, SM O'Toole, GB Keller, Nmda receptors in visual cortex are necessary
  for normal visuomotor integration and skill learning.
\newblock {\em\protect\JournalTitle{eLife}} \textbf{11}, e71476 (2020).

\bibitem{Ames2019}
KC Ames, MM Churchland, Motor cortex signals for each arm are mixed across
  hemispheres and neurons yet partitioned within the population response.
\newblock {\em\protect\JournalTitle{eLife}} \textbf{8} (2019).

\bibitem{li2009}
F Li, JZ Tsien, Memory and the nmda receptors.
\newblock {\em\protect\JournalTitle{N Engl J Med.}} \textbf{16}, 302--303
  (2009).

\end{thebibliography}

\end{document}